\documentclass[jounral]{IEEEtran}
\IEEEoverridecommandlockouts
\usepackage{cite}
\usepackage{graphicx}
\usepackage{epstopdf}
\usepackage{amsmath,amssymb,amsfonts}
\usepackage{algorithmic}
\usepackage{textcomp}
\usepackage{xcolor}
\usepackage{pifont}
\newcommand{\gcheckmark}{\textcolor{green}{\ding{51}}} 
\newcommand{\rcrossmark}{\textcolor{red}{\ding{55}}}   
\usepackage{colortbl}
\usepackage{xcolor}
\usepackage{tikz}
\usetikzlibrary{arrows.meta, positioning, decorations.markings}
\usepackage{tabularx,booktabs}
\usepackage[flushleft]{threeparttable}

\usepackage{booktabs,ragged2e}
\usepackage{multirow}
\usepackage{caption}
\usepackage{subcaption}

\usepackage{bm}
\usepackage[colorlinks,
            linkcolor=blue,
            anchorcolor=blue,
            citecolor=blue
            ]{hyperref}

\def\BibTeX{{\rm B\kern-.05em{\sc i\kern-.025em b}\kern-.08em
    T\kern-.1667em\lower.7ex\hbox{E}\kern-.125emX}}

\begin{document}
\bstctlcite{IEEEexample:BSTcontrol}
\title{Retrieval-augmented Generation for GenAI-enabled Semantic Communications}

\author{
Shunpu Tang,
Ruichen Zhang, 
Yuxuan Yan, 
Qianqian Yang, \\
Dusit Niyato, ~\IEEEmembership{Fellow,~IEEE},
Xianbin Wang, ~\IEEEmembership{Fellow,~IEEE},
and Shiwen Mao, ~\IEEEmembership{Fellow,~IEEE}

 \thanks{S. Tang, Y. Yan, and Q. Yang are with the College of Information Science and Electronic Engineering, Zhejiang University, Hangzhou, China (email: \{tangshunpu, yanyx44, qianqianyang20\}@zju.edu.cn).}
 \thanks{R. Zhang and D. Niyato are with the College of Computing and
Data Science, Nanyang Technological University, Singapore (e-mail:
\{ruichen.zhang, dniyato\}@ntu.edu.sg).}
\thanks{X. Wang is with the Department of Electrical and Computer Engineering, Western University, London, ON N6A 5B9, Canada
(e-mails: xianbin.wang@uwo.ca)}
\thanks{S. Mao is with the Department of Electrical and Computer Engineering, Auburn University, Auburn, AL 36849 USA (e-mail: smao@ieee.org).}
}

\maketitle

\thispagestyle{empty}
\pagestyle{empty}
\begin{abstract}
Semantic communication (SemCom) is an emerging paradigm aiming at transmitting only task-relevant semantic information to the receiver, which can significantly improve communication efficiency. Recent advancements in generative artificial intelligence (GenAI) have empowered GenAI-enabled SemCom (GenSemCom) to further  expand its potential in various applications. However, current GenSemCom systems still face challenges such as semantic inconsistency,  limited adaptability to diverse tasks and dynamic environments, and the inability to leverage insights from past transmission. Motivated by the success of retrieval-augmented generation (RAG) in the domain of GenAI, this paper explores the integration of RAG in GenSemCom systems. Specifically, we first provide a comprehensive review of existing GenSemCom systems and the fundamentals of RAG techniques. We then discuss how RAG can be integrated into GenSemCom. Following this, we conduct a case study on semantic image transmission using an RAG-enabled diffusion-based SemCom system, demonstrating the effectiveness of the proposed integration. Finally, we outline future directions for advancing RAG-enabled GenSemCom systems.

\end{abstract}

\begin{IEEEkeywords}
Semantic communication, Generative artificial intelligence, Retrieval-augmented generation, LLM, Diffusion model.
\end{IEEEkeywords}

\section{Introduction}

Semantic communication (SemCom) \textcolor{black}{has emerged as a promising approach for the upcoming 6G network}. Unlike traditional digital communications that focus on \textcolor{black}{reliably transmitting bits through noisy channels without considering their underlying meaning}, SemCom aims to transmit only the most important and relevant semantic information that benefits the task of the receiver\cite{Semantic_survey}. This task-oriented approach significantly improves communication efficiency and robustness, particularly \textcolor{black}{in challenging channel conditions with noise and interference, which addresses the increasing demands for data transmission in future applications.}

To realize this vision, researchers have explored how artificial intelligence (AI) technologies can enable the extraction, transmission, and reconstruction of semantic information\cite{Jiacheng_SemCom}. Numerous AI-based SemCom systems and related applications have been developed, particularly those that take advantage of discriminative models. However, existing solutions \textcolor{black}{face limitations due to the constrained model capacity} and discriminative learning paradigms, struggling to achieve both extreme bandwidth efficiency and semantic fidelity during transmission\cite{Shunpu_SemCom}.

Fortunately, generative AI (GenAI) presents new opportunities for the SemCom systems due to its remarkable capabilities in content generation, understanding, and transformation\cite{xia2023generative}. In particular, large language models (LLMs) and generative diffusion models (GDMs) approximate the underlying probability distribution of data by going beyond the extraction of discriminative features, which facilitates GenAI-enabled SemCom (GenSemCom) to reconstruct the original content with high semantic consistency at the receiver. In this direction, existing studies have been proposed to explore the potential of GenSemCom systems in various application contexts\cite{GAN_JSCC}. Despite these advancements, GenSemCom still faces several critical  challenges outlined as follows:

\begin{itemize}
    \item \textbf{Semantic inconsistency under challenging conditions.} GenSemCom systems may fail to maintain semantic consistency under limited bandwidth or severe channel noise. This is because the GenAI models used in these systems \textcolor{black}{may produce} hallucinated or fabricated output when exposed to insufficient, ambiguous, or distorted prompts\footnote{\url{https://huggingface.co/papers/2406.09358}}. 
    
    \item \textbf{Limited adaptability to dynamic tasks and environments.} GenSemCom often struggles to adapt to diverse domains, dynamic environments, and varying task requirements. \textcolor{black}{This reason is that} GenAI models typically trained on static datasets that may not generalize well to new and unseen scenarios or out-of-domain inputs\footnote{\url{https://openreview.net/forum?id=KTrnOhAN4k}}. 
    
    \item \textbf{Lack of knowledge accumulation.} GenSemCom systems are unable to leverage historical transmission data to improve performance. \textcolor{black}{This is due to the fact that} they lack mechanisms for effectively storing past interactions and leveraging them to guide and refine the generation process of GenAI models\cite{tang2024evolving}.  
\end{itemize}

\definecolor{red_GenSemCom}{RGB}{238,125,49}
\definecolor{blue_GenSemCom}{RGB}{78,149,217}
\begin{figure*}
    \centering
    \includegraphics[width=0.98\linewidth]{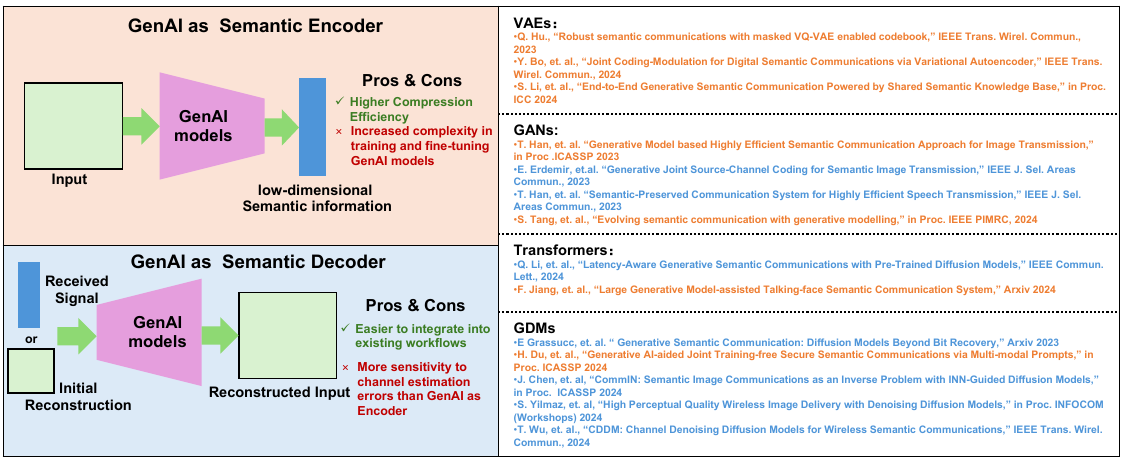}
    \caption{Overview of the representative works about GenSemCom in the past two years (2023-2024), where \textcolor{red_GenSemCom}{yellow} and \textcolor{blue_GenSemCom}{blue} are used to denote the works mainly focusing on using GenAI as the semantic encoder and decoder, respectively. }
    \label{fig:GenSemCom_overview}
\end{figure*}

To address these challenges, \textcolor{black}{incorporating retrieval-augmented generation (RAG) into GenSemCom presents a promising solution. RAG introduces retrieval mechanisms into GenAI models to dynamically access and integrate external knowledge or historical data. \textcolor{black}{This can bring} the potential to generate more accurate, context-aware, and adaptable outputs in diverse and dynamic scenarios\footnote{\url{https://github.com/NirDiamant/RAG_Techniques}}.} Motivated by these advantages, this paper aims to analyze and discuss how RAG can enhance the performance of GenSemCom systems. Specifically, we begin with a brief overview of GenSemCom systems, followed by an introduction to the fundamentals of RAG. We then present the architecture of the proposed RAG-enabled GenSemCom system and discuss its potential benefits. Finally, we conduct a case study focusing on image transmission, which demonstrates the superior reconstruction performance of our proposed system compared to existing GenSemCom systems. The main contributions of this paper can be summarized as follows: 

\begin{itemize}
    \item We \textcolor{black}{comprehensively review} the existing GenSemCom systems and discuss the two primary approaches to integrate GenAI models into SemCom systems. We also provide \textcolor{black}{an extensive} introduction to RAG, including its foundational concepts, operational principles across different GenAI models, and real-world applications.
    
    \item We propose a novel RAG-enabled GenSemCom system, comprising a knowledge base, an intelligent retriever, a knowledge-aware semantic encoder, and a decoder. \textcolor{black}{We provide insights into the design and construction of this \textcolor{black}{components}, analyze their functions in the SemCom process, and outline the system's workflow and key benefits.}

    \item We conduct a case study on semantic image transmission by integrating RAG into a GDM-based GenSemCom system. We evaluate our proposed system \textcolor{black}{based on} various metrics and the numerical results demonstrate that RAG \textcolor{black}{can significantly enhance} the performance of GenSemCom, achieving superior image fidelity and semantic consistency compared to existing  GenSemCom systems. 
\end{itemize}


\begin{figure*}[!ht]
    \centering
   \includegraphics[width=0.94\textwidth]{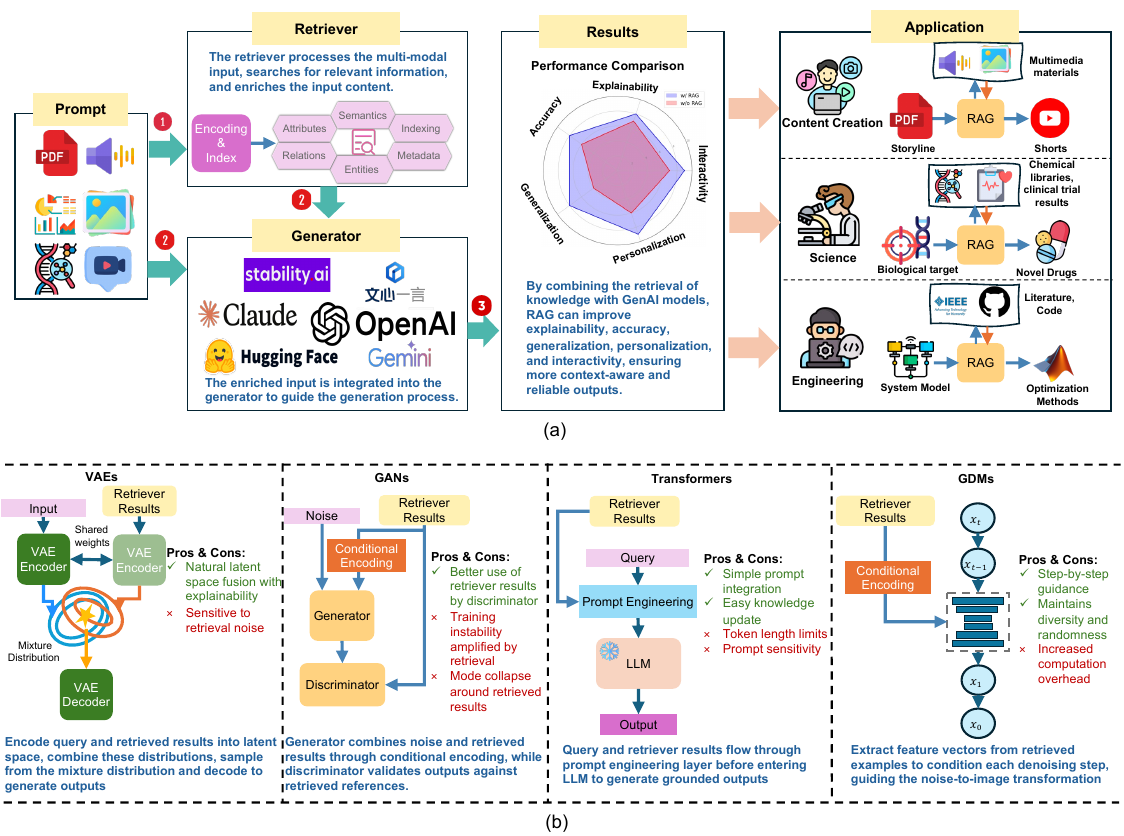}
    \caption{Illustration of RAG. (a) The components and the typical procedure of RAG, including inputted prompt, retriever, and generator. (b) Different ways of integrating RAG in GenAI models including VAEs, GANs, transformers, and GDMs.}
    \label{fig:RAG}
\end{figure*}
\section{Overview of GenSemCom and RAG}
In this section, we first briefly overview GenSemCom with different GenAI models. We then introduce the fundamentals of RAG, including its core components, and discuss how to integrate it within various GenAI workflows.

\subsection{Overview of GenSemCom}
GenSemCom represents a novel paradigm shift from existing discriminative AI-based SemCom systems by leveraging the powerful generative capabilities of GenAI models, such as VAEs, GANs, Transformers, and GDMs, to achieve efficient semantic extraction, compression, transmission, and reconstruction. \textcolor{black}{As shown in \autoref{fig:GenSemCom_overview}, we \textcolor{black}{present recent} representative works that explore integrating GenAI into SemCom systems:}


\subsubsection{\textbf{GenAI as Semantic Encoder}}
 GenAI models can understand the transmitted information and map it into a low-dimensional representation. Specifically, we can use the reverse process of GenAI models such as GANs and GDMs. The forward process is to sample from a noise sample and refine it into meaningful data, such as images, text, and audio. Then, the reverse process aims to map a given input back into a corresponding noise representation. Therefore, this noise representation can serve as the semantic information of the input data. At the receiver, the original input can be reconstructed through the forward generation process of GenAI models. For example, the authors in \cite{tang2024evolving} proposed a channel-aware GAN inversion approach to extract the feature-disentangled and channel-correlated latent code of transmitted images that can be directly transmitted over noisy channels.

\subsubsection{\textbf{GenAI as Semantic Decoder}}
GenAI models can \textcolor{black}{support} semantic decoding by reconstructing transmitted data with high fidelity and semantic consistency. Since the goal of GenAI models is to learn and approximate the original data distribution, \textcolor{black}{the models} can sample from this distribution conditionally to generate outputs that closely resemble the input data according to the received signals. For example, the authors in \cite{GAN_JSCC} treated the wireless transmission process as a degradation process, and exerted an estimated degradation process to the generated samples. By minimizing the difference between the degraded version of the generated samples and the received degraded data, the generated data will converge to approximate the original one.


\textcolor{black}{However, existing GenSemCom systems face several challenges, including the generation of hallucinated content, limited generalization across diverse data distribution and environments, and the inability to accumulate and leverage communication history. These limitations motivate us to integrate RAG and further explore the untapped potential of GenSemCom systems.}


\subsection{Overview of \textcolor{black}{Retrieval-augmented Generation}}

A typical RAG process is shown in \hyperref[fig:RAG]{\autoref{fig:RAG}(a)}. After the user inputs the prompt, the retriever will extract the key feature of the prompt and query the knowledge base or database for relevant information. The retrieved information is then combined with the original prompt and will be used by the generator to produce the desired context-aware content. Specifically, there are two types of retrievers in RAG as follows:
\begin{itemize}
    \item \textbf{Sparse Retriever (SR)}: SR approaches are widely used in the domain of text and document retrieval, where sparse representations, such as keywords and word frequency, are employed to identify relevant documents. Key technologies in this area include the term frequency-inverse document frequency (TF-IDF)\footnote{\url{https://github.com/mayank408/TFIDF}}, BM25\footnote{\url{https://github.com/dorianbrown/rank_bm25?tab=readme-ov-file}}, and inverted index\footnote{\url{https://www-cs-faculty.stanford.edu/~knuth/taocp.html}}. \textcolor{black}{Due to the nature of sparse representations, an SR can efficiently handle large text volumes in scenarios such as search engines and large-scale retrieval systems where interpretability, scalability, and response latency are critical. However, it often struggles to capture deeper semantic relationships and contextual nuances within the text, limiting its effectiveness in tasks that require advanced semantic understanding, fuzzy matching or handling multimodal data.} 

    \item \textbf{Dense Retriever (DR)}: DR approaches can retrieve relevant information based on the semantic similarity of the dense representation such as embeddings and features generated from the input data using a pre-trained deep neural network, such as BERT\footnote{\url{https://github.com/google-research/bert}} and ViT\footnote{\url{https://github.com/google-research/vision_transformer}} models. \textcolor{black}{Compared with SR, a DR provides superior semantic or fuzzy retrieval performance and enables cross-modal retrieval through contrastive learning\footnote{\url{https://openai.com/index/clip/}}. These make the DR highly suitable for applications such as question-answering and recommendation systems.}
\end{itemize}

Moreover, since different GenAI models exhibit varying architectures and pipelines, as shown in \hyperref[fig:RAG]{\autoref{fig:RAG}(b)}, we provide an overview of the typical methods for integrating RAG into different GenAI models:
\begin{itemize}
    \item \textbf{RAG for VAE}: RAG can be incorporated with VAE by retrieving external information and encoding the retrieval results into the latent space using the VAE encoder. By mixing the distribution of original input and retrieval results and sampling from the new distribution, the extra information can be used to improve the generated outputs during the VAE decoding process\cite{deng2023regavae}, while still maintaining the probabilistic nature of the original VAE. However, the approach of mixing multiple distributions is sensitive to retrieval noise.
    
    \item{\textbf{RAG for GANs}:} RAG can enhance GANs by encoding retrieved results and using them as additional conditioning inputs for the GAN generator, which can guide the generator to produce more realistic and contextually relevant outputs. For example, the authors in \cite{casanova2021instance} proposed Instance-Conditioned GAN (IC-GAN), which takes both a noise vector and retrieved instance features as inputs to model a neighborhood distribution around the instance. The discriminator then distinguishes between generated and real samples drawn from the neighborhood of that instance. However, similar to conventional GANs, RAG-enhance GANs still suffer from instability during training, especially when extra conditions exist.
    
    \item{\textbf{RAG for Transformer}:} RAG can introduce external results to enhance the GenAI models based on Transformer, such as LLMs. A typical process is demonstrated in \cite{zhao2024towards}, the query is processed by a retriever that is used to fetch the most relevant documents from a large dataset. These retrieved documents are then integrated with the original query using prompt engineering techniques. Therefore, the model can generate more contextually aware responses according to extra knowledge, and effectively reduce the risk of hallucinations. However, such solutions may face challenges, as the maximum token limits of transformer models prevent them from processing large inputs, leading to truncation of external knowledge.

    \item{\textbf{RAG for GDMs}:} RAG can also enhance GDMs by retrieving relevant reference images or styles and encoding them as condition guidance during the denoise steps.  For example, the work in \cite{chen2022re} enhances diffusion-based text-to-image generation by providing high-level semantic understanding and low-level visual details about the target entities, and enables the GDM to produce images aligned well with user intentions and preferences, even when the object is unseen in the training dataset. However, GDMs are inherently computationally intensive due to their iterative denoising process, and the integration of RAG further increases inference time and computational demands.
\end{itemize}

\subsection{Application of RAG}
Thanks to RAG's ability to integrate retrieval mechanisms with GenAI models, its applications extend beyond traditional text, audio, and image generation to fields such as scientific research and engineering design. Specifically, in drug discovery, protein structure prediction, and molecular generation, RAG can provide valuable guidance by retrieving domain-specific knowledge to enhance the generative process. For example, the authors in \cite{DBLP:conf/iclr/0001NQXBA23} proposed incorporating exemplar molecules that satisfy specific design constraints into the generative pipeline. By leveraging these exemplars as retrieval-enhanced inputs, the generative model can produce novel molecules that adhere to desired properties. 

Moreover, in engineering domains such as networking and wireless communication, RAG can facilitate the optimization of resource allocation,  network protocol design, and performance prediction by retrieving relevant configurations, cases, and 
system implementation. For example, the authors in \cite{Ruichen_RAG} demonstrated the use of LLMs enhanced by RAG to address optimization problems in satellite communication, where relevant satellite expert knowledge is retrieved to facilitate effective mathematical modeling and decision-making processes.

\textbf{Lesson Learned:} From the success of RAG in these applications, it is evident that integrating RAG with GenAI models has the potential to revolutionize various fields. By combining the retrieval of domain-specific knowledge with the generative capabilities of GenAI, RAG not only enhances the contextual accuracy of generated outputs but also extends the applicability of GenAI models to complex tasks, which motivated us to explore the integration of RAG into GenSemCom systems, aiming to further improve the semantic understanding, transmission efficiency, and robustness of GenSemCom systems.

 \section{RAG-enabled GenSemCom}
\begin{figure*}
     \centering
     \includegraphics[width=0.9\textwidth]{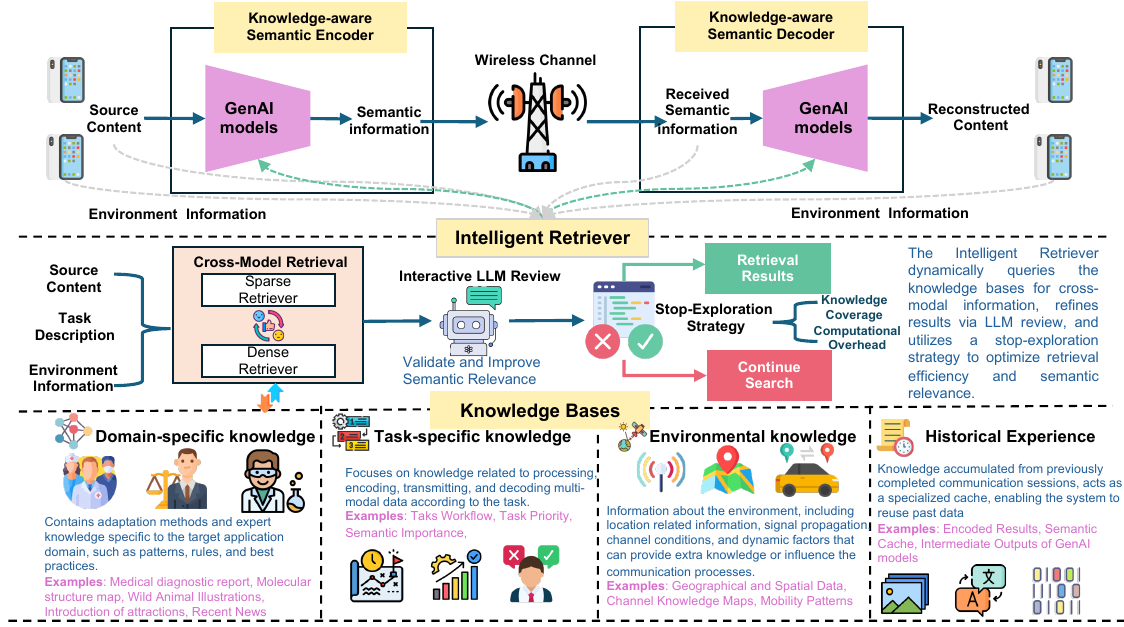}
     \caption{Illustration of the proposed RAG-enabled GenSemCom system, where the intelligent retriever dynamically queries knowledge bases and the retrieved results are refined through interactive LLM reviews. Besides, a stop-exploration strategy is used to balance efficiency and relevance. The knowledge-aware semantic encoder and decoder use the retrieved information to transmit and reconstruct the content with high semantic consistency.}
     \label{fig:framework}
 \end{figure*}
 
 In this section, we elaborate how RAG can be incorporated into GenSemCom to enhance system performance from different perspectives. First, we introduce the key components of RAG-enabled GenSemCom. Then we give an overall overflow of RAG-enabled GenSemCom.
 \subsection{Components of RAG-enabled GenSemCom}
Unlike traditional GenSemCom, where GenAI models rely on knowledge obtained from pre-training on large datasets, RAG-enabled GenSemCom can dynamically retrieve relevant additional knowledge to enhance the semantic encoding or decoding processes. As shown in \autoref{fig:framework}, the key components and workflow of the RAG-enabled GenSemCom \textcolor{black}{are included as follows:}
\subsubsection{\textbf{Knowledge base}} \textcolor{black}{The knowledge base is an external repository that includes domain-specific, task-specific, and environmental-specific knowledge and communication history, which the system dynamically retrieves to enhance the SemCom process. Specifically, domain knowledge, such as expert insights, patterns, and rules, can help GenSemCom address issues like hallucinations by providing contextually relevant information to guide the generation process.  In addition, expert insights and domain-specific adaptation methods can also improve system adaptability when handling unseen data distributions. Moreover,  task knowledge focuses on execution strategies, such as task decomposition strategy, task priority, and semantic importance, which can be used to help GenSemCom handle diverse tasks.  Environmental knowledge, such as channel knowledge maps (CKMs), offers critical insights into channel state and signal conditions, improving the robustness of semantic encoding and decoding under varying network environments\cite{ren2024knowledge}. A cache space can also be deployed in GenSemCom to store past transmission results, including encoding results, and intermediate outputs of GenAI models, which allow the system to retrieve and reuse communication history to reduce bandwidth and improve reconstruction performance of original data\cite{tang2024evolving}.}

\subsubsection{\textbf{Intelligent retriever}} \textcolor{black}{The intelligent retriever can bridge the gap between external knowledge and the generation process by dynamically querying the knowledge base to provide relevant information. It also ensures that the retrieved content aligns with the task requirements and user intent. Specifically, the retriever can leverage advanced retrieval techniques that support cross-modal and semantic retrieval to access the knowledge base. For example, one can use textual queries to retrieve relevant visual data. Second, an interactive LLM can be integrated as a reviewer to refine and enhance the retrieval results. By reasoning over the retrieved knowledge, the LLM filters out irrelevant or redundant information. This ensures that only semantically relevant content is passed to the GenAI-based semantic encoder or decoder. Moreover, a stop-exploration strategy\footnote{\url{https://github.com/thunlp/Adaptive-Note}} can be employed in this step to decide whether the retrieval results are sufficiently informative for subsequent generation or if further retrieval is necessary to gather more information. This approach can balance knowledge exploration and computational overhead.}

\subsubsection{\textbf{Knowledge-Aware Semantic Encoder and Decoder}}
\textcolor{black}{The knowledge-aware semantic encoder and decoder are designed to enhance the semantic encoding and decoding processes. At the transmitter, the semantic encoder incorporates retrieved knowledge, including domain, task, environmental, and historical information, to refine the encoding process to better understand, compress, and encode the source data. At the receiver, the semantic decoder combines both the received signal with retrieved knowledge to reconstruct the transmitted input with high semantic consistency. To achieve this integration, we can use low-rank adaptation\footnote{\url{https://huggingface.co/docs/diffusers/en/training/lora}} to fine-tune the existing GenAI semantic encoder and decoder with retrieved knowledge or introduce extra adapters\footnote{\url{https://huggingface.co/docs/hub/adapters}} to preserve the original pre-trained weights while integrating new information. Besides, for text transmission tasks facilitated by LLMs, prompt engineering\cite{Ruichen_RAG} can effectively incorporate retrieved knowledge into the encoding and decoding processes.}

\subsection{Overall Workflow of RAG-enabled GenSemCom}
The workflow of RAG-enabled GenSemCom is as follows. First, the retriever dynamically queries relevant knowledge from the knowledge base based on the source content and environmental data. The retrieved knowledge, refined through the interactive LLM reviewer, assists the semantic encoder in extracting semantic information for transmission. The extracted semantic information is then transmitted over wireless channels. At the receiver, the knowledge-aware semantic decoder combines the received semantic data with retrieved knowledge to reconstruct the original content with high semantic consistency. Furthermore, the system updates the knowledge base by storing historical transmission data, facilitating future reuse and continuous improvement.

\begin{figure*}
   \centering
        \includegraphics[width=0.68\textwidth]{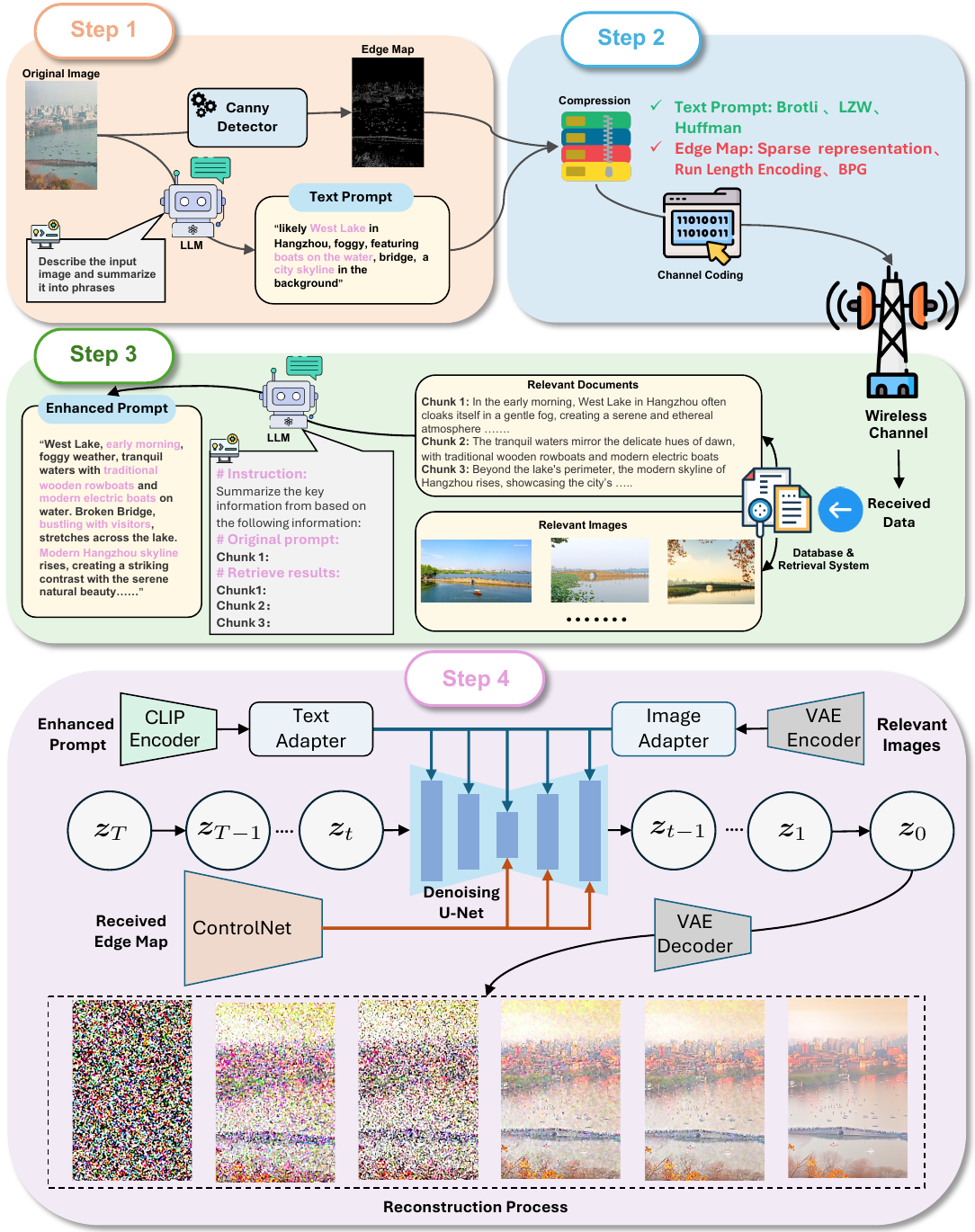}
        \caption{Illustration of the proposed GDM-based SemCom system with RAG. Key steps include: (1) \textbf{Multimodal semantic information extraction} using LLMs for text prompts and Canny detector for edge maps as visual prompts; (2) \textbf{Semantic information transmission} after source coding and channel coding; (3) \textbf{Information retrieval and prompt enhancement}, retrieving related textual and visual information using RAG to enhance prompts; (4) \textbf{Image reconstruction}, using GDM with ControlNet to reconstruct high-quality images. }
        \label{fig:case_study}
\end{figure*}

\begin{figure*}[!t]
    \centering
    \begin{subfigure}[t]{0.3\textwidth}
        \centering
        \includegraphics[width=\textwidth]{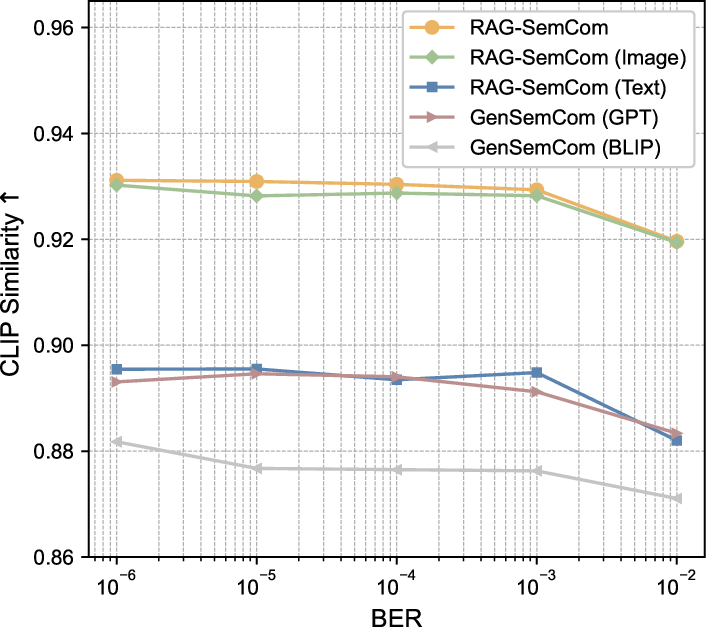}
        \caption{CLIP similarity on the Kodak dataset.}
        \label{fig:kodak}
    \end{subfigure}
    \begin{subfigure}[t]{0.3\textwidth}
        \centering
        \includegraphics[width=\textwidth]{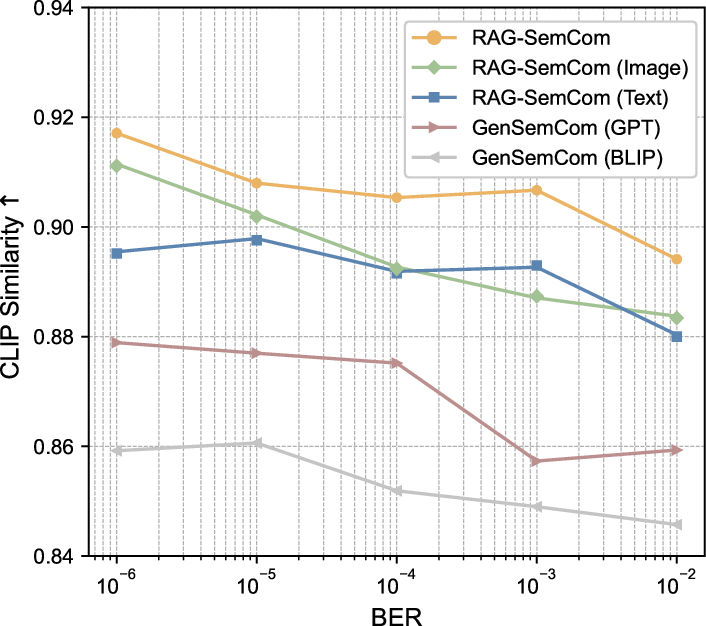}
        \caption{CLIP similarity on the West Lake image.}
        \label{fig:west_lake}
    \end{subfigure}

    \begin{subfigure}[t]{0.67\textwidth}
        \centering
        \caption{Ablation study on the West Lake image.} 
        \vspace{0.5em} 
        \label{fig:rag_results}
        \resizebox{\textwidth}{!}{
        \begin{tabular}{lcc|cccc}
            \toprule
            \multicolumn{2}{r}{\textbf{Setting}} & & \multicolumn{4}{c}{\textbf{Metric}} \\ 
            \cmidrule(lr){2-3} \cmidrule(lr){4-7}
            \textbf{Methods} & \textbf{Image RAG} & \textbf{Text RAG} & \textbf{LPIPS $\downarrow$} & \textbf{PIEAPP $\downarrow$} & \textbf{MS-SSIM $\uparrow$} & \textbf{CLIP Similarity $\uparrow$} \\
            \midrule
            GenSemCom (GPT)             & \rcrossmark        & \rcrossmark        & 0.4388                      & 3.1318                      & 0.5981                      & 0.8752                   \\
            \midrule
            RAG-enabled GenSemCom   & \rcrossmark        & \gcheckmark       & 0.4342                      & 3.1909                      & 0.6228                      & 0.8919                      \\
                       & \gcheckmark        & \rcrossmark       & \cellcolor{pink!50}0.3985   & \cellcolor{pink!50}2.1111   & \cellcolor{pink!50}0.6712   & \cellcolor{blue!25}0.8927  \\
                       & \gcheckmark        & \gcheckmark       & \cellcolor{blue!25}0.4075   & \cellcolor{blue!25}2.1841   & \cellcolor{blue!25}0.6682   & \cellcolor{pink!50}0.9053   \\
            \bottomrule
        \end{tabular}
        }
    \end{subfigure}

\caption{Numerical results of the proposed RAG system with different configurations. (a) CLIP similarity across varying BERs on the Kodak dataset. (b) CLIP similarity across varying BERs on the West Lake image. (c) Ablation study on the West Lake image.}
\label{fig:combined}
\end{figure*}
\section{Case Study: RAG-enabled GenSemCom for image transmission with Multi-modal Prompts}

In this section, we conduct a case study to explore the effectiveness of RAG-enabled GenSemCom systems and demonstrate how the incorporation of RAG can enhance the quality of image transmission.
\subsection{Proposed System}
As illustrated in \autoref{fig:case_study}, we propose a RAG-enabled GenSemcom system that uses a GDM as the semantic encoder and incorporates RAG techniques along with multi-modal prompts to improve system performance. The proposed system comprises a transmitter equipped with a multi-modal semantic encoder and a receiver with a semantic decoder incorporating an RAG system and an associated knowledge database. The semantic encoder transforms input images into a multi-modal representation and then transmits them to the receiver, where the semantic decoder uses GDMs to reconstruct the images. The procedure of the proposed system can be summarized as follows.

\subsubsection{\textbf{Multimodal semantic information extraction}}
Following \cite{GemSom_GDM},  the multimodal semantic encoding begins by processing an input image by extracting associated textual descriptions using well-trained state-of-the-art (SOTA) vision-language models or multi-modal LLMs, since the textual descriptions can be regarded as the high-level semantic information and can be used as text prompt at the receiver\cite{xia2023generative, zhang2024toward}. Moreover, to preserve critical structural details of images during transmission and mitigate the impact of potential semantic noise introduced by the textual description process, the encoder extracts the edge map of the input image, which the important geometric and boundary information can be regarded as a visual prompt to guide the denoise process of GDMs in a conditional manner.

\subsubsection{\textbf{Semantic information transmission}}
To further optimize transmission efficiency, we employ source coding strategies for multi-model semantic information. Specifically, the extracted textual descriptions undergo Brotli compression\footnote{\url{https://github.com/google/brotli}}, and the edge map can be encoded using sparse representation, run-length encoding to transmit the indices of edge pixels or be regarded as a gray image and compressed by an image compression approach, such as BPG\footnote{\url{https://bellard.org/bpg/}}. The compressed semantic information is then transmitted over noisy channels after channel coding.

\subsubsection{\textbf{Information retrieval and prompt enhancement}}

At the receiver, we first decode and obtain the textual descriptions and edge map. Next, we retrieve the related information in the knowledge base. Specifically, the textual descriptions are leveraged as queries to retrieve related and more detailed information from the knowledge base, including supplementary text and documents that enrich the semantic context. We then exert prompt engineering to combine the original text and retrieval results. This process can be facilitated by LLMs, which enhances the prompt by ensuring semantic coherence, and contextual relevance. Moreover, the RAG system retrieves related images based on the textual descriptions and the edge map. These images provide visual context and supplementary knowledge to the following generation process of GDM.

\subsubsection{\textbf{Image reconstruction}}
Finally, we use the advanced GDM equipped with ControlNet\footnote{\url{https://github.com/lllyasviel/ControlNet}} to reconstruct the image according to the optimized text prompt, the decoded edge map, and the retrieved related images. Specifically, the text prompt and the retrieved images are input to the text encoder and image encoder, respectively, and then passed to the UNet denoiser.  At the same time, the ControlNet processes the edge map to provide structural guidance by combining the putout of UNet. Through serval iterations, the system generates the final high-quality reconstructed image with semantic, contextual, and structural consistency.

\subsection{Simulation Settings}
We implement the proposed system using Pytorch on a Nvidia A6000 GPU. The edge maps are extracted using the Canny edge detector with fixed thresholds. For generating textual descriptions of the images and performing retrieval tasks, we utilize the advanced GPT-4o model\footnote{\url{https://platform.openai.com/docs/models/gpt-4o}}. For GDM, we use the pre-trained stable-diffusion XL model implemented by diffuser framework\footnote{\url{https://github.com/huggingface/diffusers}}. To enhance the multi-model prompt capability of the stable diffusion model, we integrate the IP-Adapter\footnote{\url{https://github.com/tencent-ailab/IP-Adapter}}, which improves conditioning control by providing more effective outputs from its text encoder and image encoder. To simulate the impact of fading channels, we follow the method in \cite{GemSom_GDM}, introducing bit error rates (BER) into the edge map transmission. \textcolor{black}{We evaluate the proposed system on the Kodak dataset and a complex image of West Lake using multiple metrics. The West Lake image is selected because it significantly differs from the distribution of the training dataset, making it ideal for testing the system's domain adaptation abilities. Besides, its details and rich semantic information provide a challenging benchmark to assess the robustness and generalization ability of GenSemCom. For evaluation, we assess semantic consistency using CLIP similarity, which measures the alignment of text-image embeddings. We employ LPIPS and PIEAPP to evaluate perceptual quality, reflecting human-preferred image characteristics. Additionally, we report MS-SSIM to measure structural fidelity, indicating how closely image structures and patterns match those of the ground truth.}

\subsection{Result Analysis}
\textcolor{black}{In \autoref{fig:kodak} and \autoref{fig:west_lake}, we first compare the CLIP similarity under different BER of the proposed RAG-enabled GenSemCom with the existing GenSemCom system\cite{GemSom_GDM}. To ensure a fair comparison, both methods employ the same Stable Diffusion model. Moreover, we consider two variants of GenSemCom according to \cite{GemSom_GDM}: one uses the GPT4o model, and the other uses BLIP to generate the original text prompt at the transmitter. We also present results for two variants of the proposed RAG-enabled GenSemCom: one that uses RAG only to provide relevant images for improved guidance and the other that only enhances the prompt with the retrieved documents. From \autoref{fig:kodak}, we can observe that the proposed RAG-enabled GenSemCom with both text and image retrieval consistently outperforms both the existing GenSemCom variants on the Kodak dataset. We can also find that the performance gains mainly come from using RAG to provide relevant images, while enhancing the text prompt with retrieved documents still offers improvements over the baseline GenSemCom systems. In \autoref{fig:west_lake}, while the transmitted image becomes more complex, we can observe the gain in performance remains consistent across the variations of RAG-enabled GenSemCom. Specifically, leveraging both image and text achieves significantly superior semantic consistency compared to the baseline GenSemCom systems without using RAG, even under varying BER levels. These results highlight the robustness and effectiveness of the proposed RAG-enabled GenSemCom in maintaining high semantic fidelity and reconstruction quality across diverse scenarios. }

In \autoref{fig:rag_results}, we conduct an ablation study of the RAG-enabled GenSemCom approach against existing GenSemCom systems,  with a BER of $10^{-4}$. We can observe that the proposed system with text-only retrieval shows significant improvements in CLIP similarity, indicating enhanced semantic consistency. By providing similar images as references, the proposed system achieves the best visual quality and fidelity, as evaluated by LPIPS, PIEAPP, and MS-SSIM. When both text and image prompts are enhanced, the system achieves the highest CLIP similarity. These results further highlight the effectiveness of incorporating RAG into GSemCom, which leverages additional knowledge to guide the generative process of GenAI models.

\begin{figure*}[!t]
\centering
\begin{subfigure}[t]{0.18\textwidth}
    \centering
    \includegraphics[width=\textwidth]{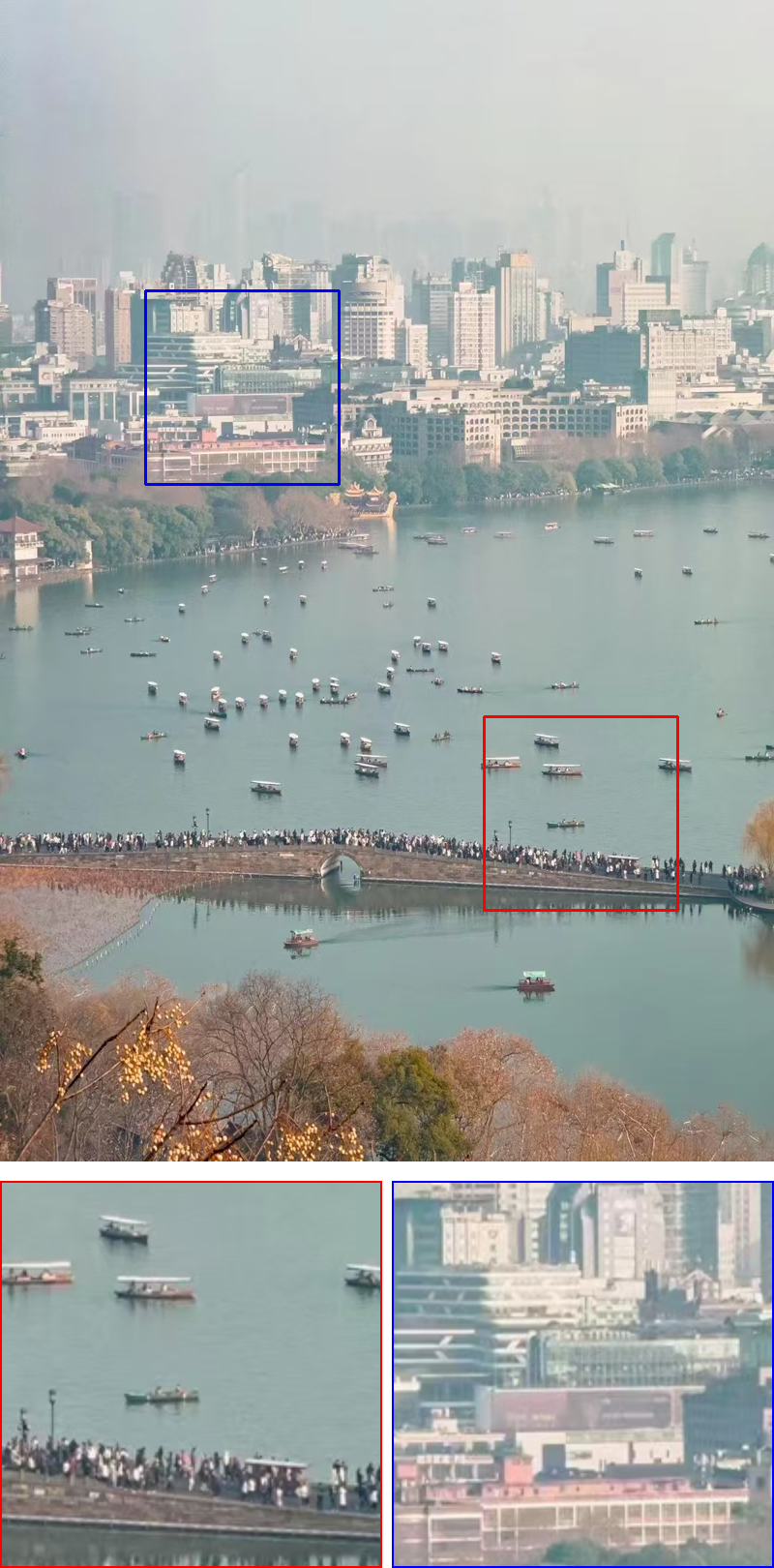}
    \caption{Original Image}
    \label{fig:original}
\end{subfigure}
\begin{subfigure}[t]{0.18\textwidth}
    \centering
    \includegraphics[width=\textwidth]{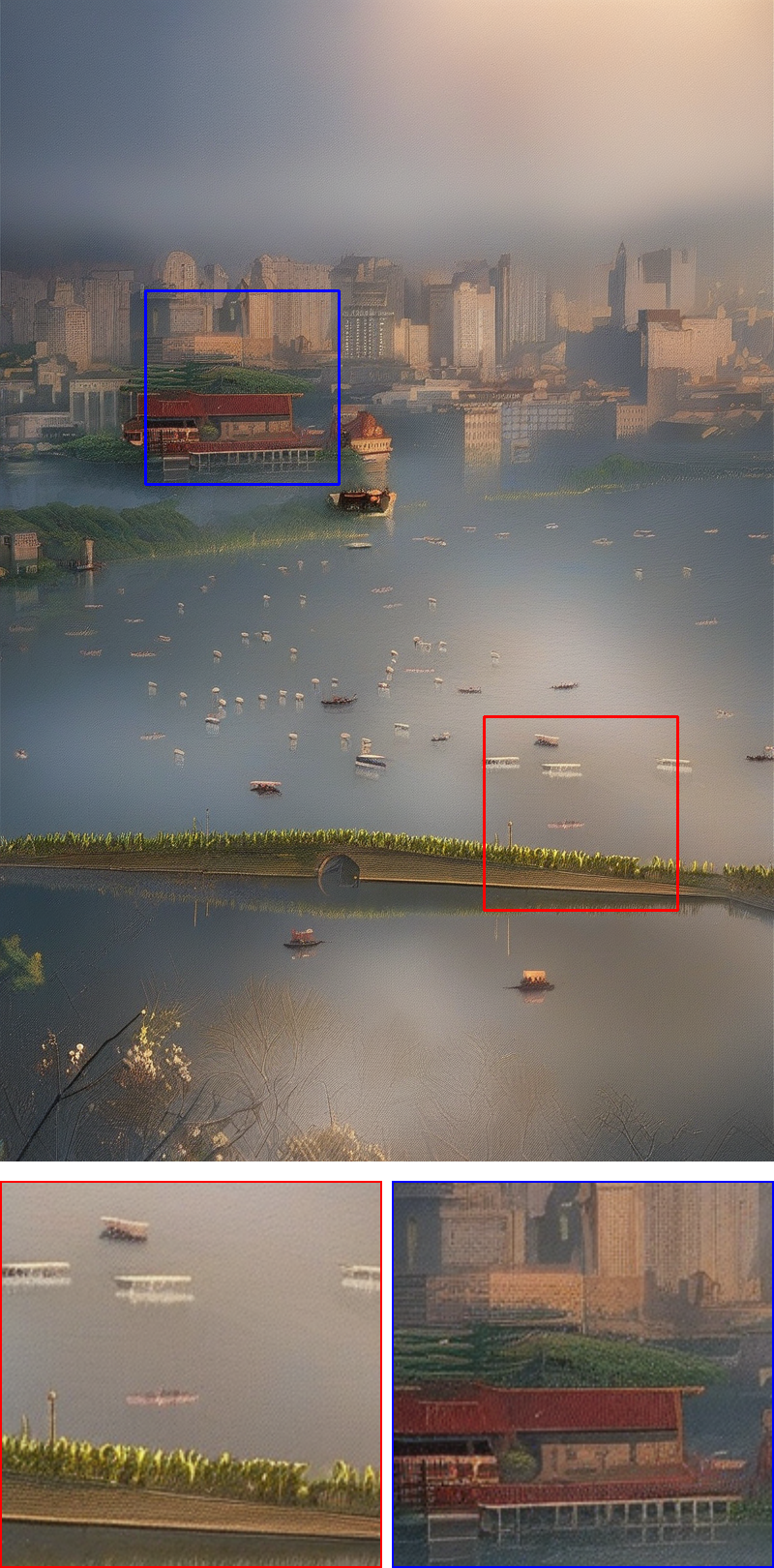}
    \caption{    \centering GenSemCom 
    (GPT)}
    \label{fig:without_rag}
\end{subfigure}
\begin{subfigure}[t]{0.18\textwidth}
    \centering
    \includegraphics[width=\textwidth]{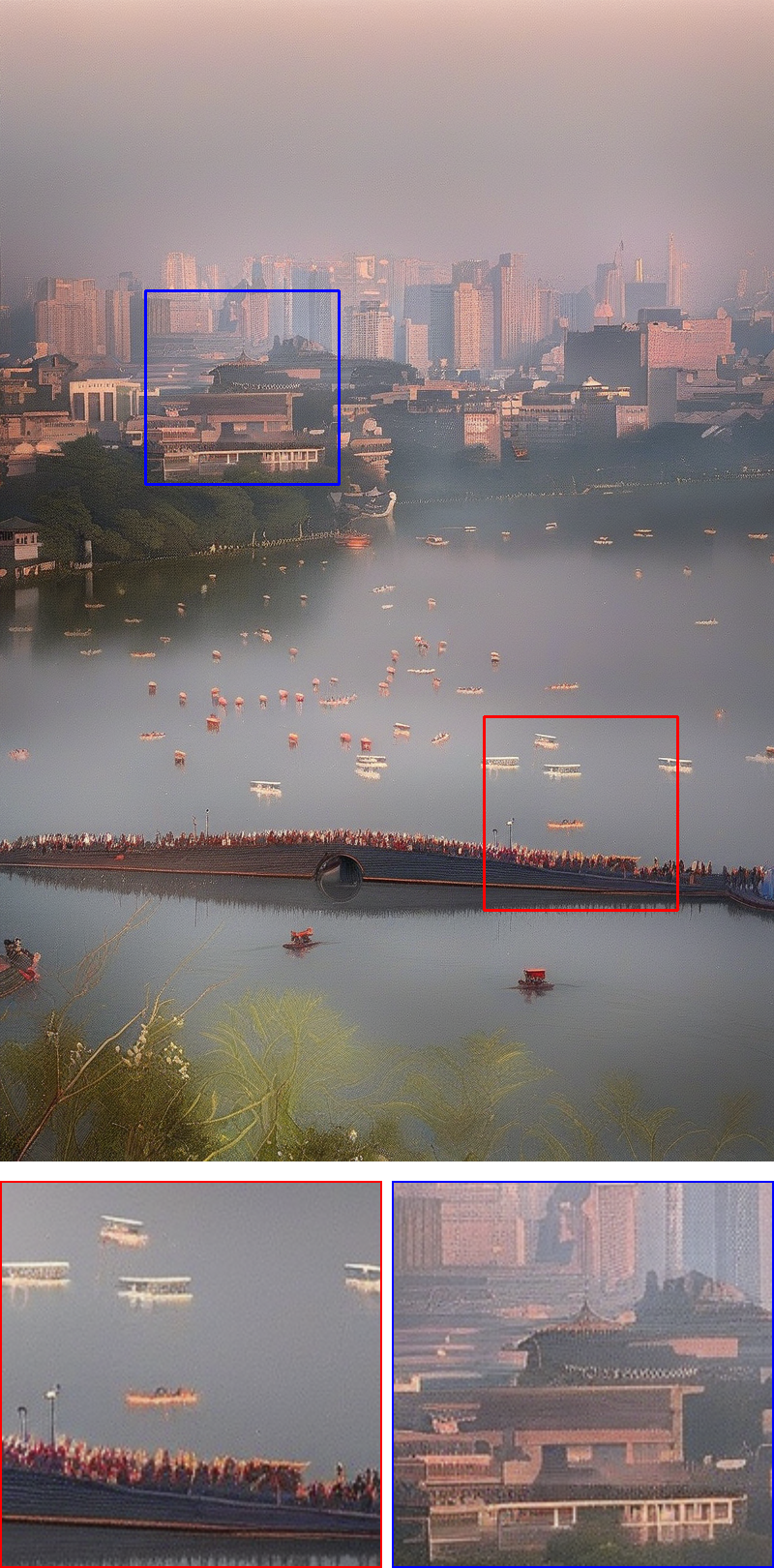}
    \caption{     \centering RAG-enabled GenSemCom (Text)}
    \label{fig:text_rag}
\end{subfigure}
\begin{subfigure}[t]{0.18\textwidth}
    \centering
    \includegraphics[width=\textwidth]{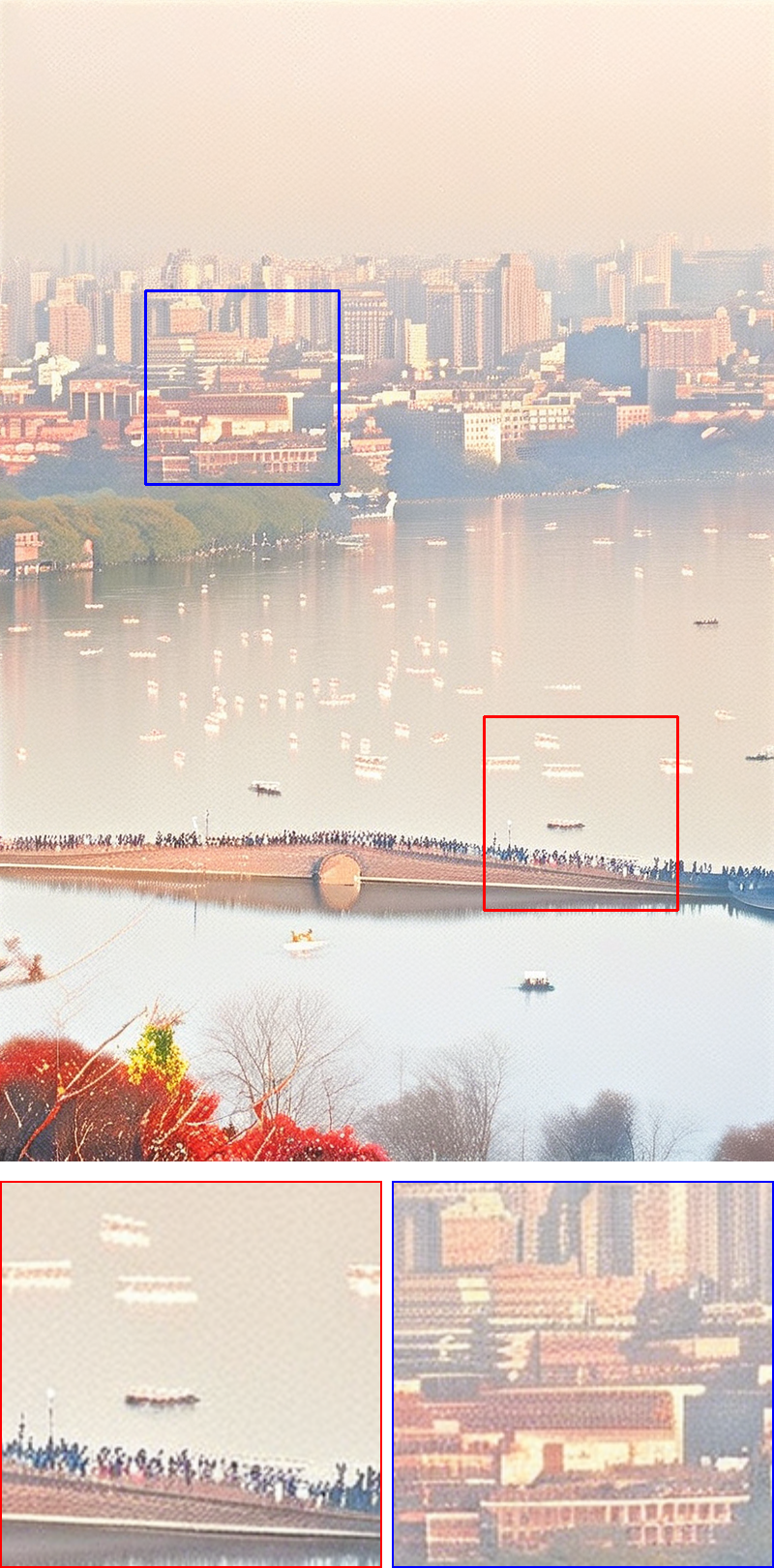}
    \caption{    \centering RAG-enabled GenSemCom (Image)}
    \label{fig:image_rag}
\end{subfigure}
\begin{subfigure}[t]{0.18\textwidth}
    \centering
    \includegraphics[width=\textwidth]{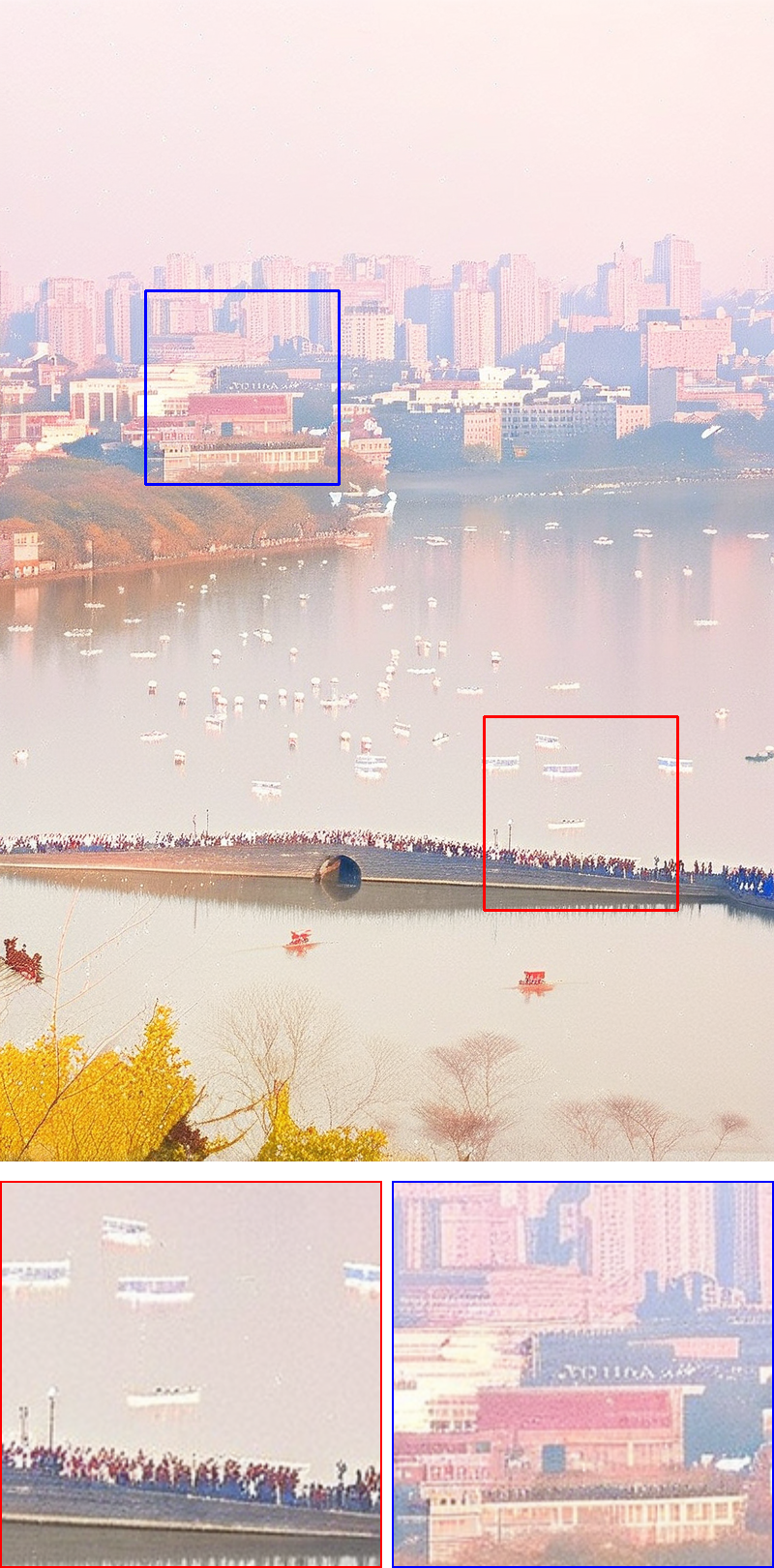}
    \caption{\centering RAG-enabled GenSemCom}
    \label{fig:full_rag}
\end{subfigure}
\caption{Comparison of visual results under different RAG configurations, where the BER is set to $10^{-3}$.}
\label{fig:rag_comparison}
\end{figure*}

We also provide visualization results as shown in \autoref{fig:rag_comparison}, where the BER is set to $10^{-3}$. From these visualizations, it is evident that the GenSemCom system may not understand the information from the edge map and simple textual prompt. This is reflected in the reconstructed image, where people on the bridge are incorrectly replaced with trees. In contrast, the proposed system can effectively overcome this issue by using the retrieved information that describes the proper semantics and details about the original image. This additional context helps the system avoid hallucinations and produce a more semantically accurate reconstruction.

\section{Future Direction}
While this work highlights the advantages of RAG-enabled GenSemCom, further advancements in this field are necessary.
\textbf{Retrieval quality and efficiency:}
To further enhance the overall effectiveness of GenSemCom, achieving the balance between retrieval accuracy and computational overhead is critical, especially in large-scale and multimodal GenSemCom systems. Future research should further explore adaptive retrieval techniques that can dynamically adjust according to task requirements.

\textbf{Synchronization of knowledge bases:} Efficiently synchronizing knowledge bases across devices, remains a challenge, as the volume of knowledge bases can be very large. Therefore, solutions such as distributed systems, federated learning, and dynamic updating protocols should be prioritized to address these issues. Moreover, it is important to develop approaches that maintain semantic consistency even in scenarios where the knowledge bases between the transmitter and the receiver are different.

\textbf{Security and privacy:}
The deployment of knowledge bases on edge networks introduces significant security and privacy concerns, including risks of privacy leakage and poisoning attacks. To mitigate these issues, future work should explore techniques in RAG-enabled SemCom such as encrypted retrieval queries, secure multi-party computation, and differential privacy mechanisms. In addition, anomaly detection approaches should be developed to identify poisoning attacks on knowledge bases.
\section{Conclusion}

In this paper, we have explored the integration of RAG with GenSemCom. We first overviewed the existing works about GenSemCom and introduced the concept and implementation of RAG in kinds of GenAI models. We then proposed a novel approach that introduced RAG into GenSemCom and highlighted the potential enhancements offered by this integration. Moreover, we conducted a case study on image transmission using RAG in a diffusion-based GenSemCom system. The evaluation across various metrics demonstrated the superiority of the proposed system over existing GenSemCom approaches. Finally, we discussed the future direction of research from several perspectives.

\bibliographystyle{IEEEtran}
\bibliography{IEEEabrv, references}

\end{document}